\documentclass[preprint,hidelinks]{elsarticle}

\usepackage{amsmath,amssymb,amsfonts,amsthm}
\usepackage{stmaryrd}
\usepackage{url}
\usepackage[table,xcdraw]{xcolor}
\usepackage{graphicx}
\usepackage{tikz}

\usepackage[hidelinks]{hyperref}

\usepackage[caption=false,font=footnotesize]{subfig}

\usepackage[acronym]{glossaries}
\glsdisablehyper
\newacronym{bpm}{BPM}{Business Process Management}
\newacronym{bpmn}{BPMN}{Business Process Model and Notation}
\newacronym{dmn}{DMN}{Decision Model and Notation}
\newacronym[plural=BPMS,firstplural=Business Process Management Systems (BPMS)]{bpms}{BPMS}{Business Process Management System}
\newacronym{omg}{OMG}{Object Management Group}
\newacronym{csl}{CSL}{Contract Specification Language}
\newacronym{mof}{MOF}{Meta\-/Object Facility}
\newacronym[plural=BCLs,firstplural=Business Collaboration Languages (BCLs)]{bcl}{BCL}{Business Collaboration Language}
\newacronym{evm}{EVM}{Ethereum Virtual Machine}
\newacronym{rim}{RIM}{Railway Infrastructure Manager}
\newacronym{erp}{ERP}{Enterprise Resource Planning}
\newacronym[plural=DApps,firstplural=Decentralized Applications (DApps)]{dapp}{DApp}{Decentralized Application}
\newacronym[plural=FSMs,firstplural=Finite State Machines (FSMs)]{fsm}{FSM}{Finite State Machine}
\newacronym{uml}{UML}{Unified Modeling Language}
\newacronym{ocl}{OCL}{Object Constraint Language}
\newacronym{dlt}{DLT}{Distributed Ledger Technology}
\newacronym[plural=DLs,firstplural=Distributed Legders (DLs)]{dl}{DL}{Distributed Ledger}
\newacronym{soa}{SOA}{Service\-/Oriented Architecture}
\newacronym{sla}{SLA}{Service Level Agreement}
\newacronym{cdm}{CDM}{Choreography Data Model} 
\newacronym{mep}{MEP}{Message Exchange Pattern}
\newacronym{rbo}{RBO}{Rental Bond Online}
\newacronym{gta}{GTA}{Grain Trade Australia}
\newacronym{xml}{XML}{Extensible Markup Language}
\newacronym{mdsd}{MDSD}{Model\-/Driven Software Design}
\newacronym{xmi}{XMI}{XML Metadata Interchange}
\newacronym{fol}{FOL}{First\-/Order Logic}
\newacronym{ufo}{UFO}{Unified Foundational Ontology}
\newacronym{ufol}{UFO-L}{Unified Foundational Ontology for Legal Relations}
\newacronym{abi}{ABI}{Application Binary Interface}
\newacronym{iata}{IATA}{International Air Transport Association}
\newacronym[plural=DSLs,firstplural=Domain\-/Specific Languages (DSLs)]{dsl}{DSL}{Domain\-/Specific Language}
\newacronym[plural=ACs,firstplural=Active Choreographies (ACs)]{ac}{AC}{Active Choreography}
\newacronym[plural=PCs,firstplural=Passive Choreographies (PCs)]{pc}{PC}{Passive Choreography}
\newacronym[plural=zkSNARKs,firstplural=Zero-Knowledge Succinct Non-Interactive Arguments of Knowledge (zkSNARKs)]{zksnark}{zkSNARK}{Zero-Knowledge Succinct Non-Interactive Argument of Knowledge}

\newtheorem{definition}{Definition}

\usepackage{xspace}
\newcommand{\storage}{storage\xspace}
\newcommand{\request}{request\-/response\xspace}
\newcommand{\req}{req/res\xspace}
\newcommand{\history}{history\xspace}
\newcommand{\publish}{publish\-/subscribe\xspace}

\newcommand{\Storage}{Storage\xspace}
\newcommand{\Request}{Request\-/response\xspace}
\newcommand{\Req}{Req/res\xspace}
\newcommand{\History}{History\xspace}
\newcommand{\Publish}{Publish\-/subscribe\xspace}
\newcommand{\Pub}{Pub/sub\xspace}
\newcommand{\async}{asynchronous\xspace}
\newcommand{\Async}{Asynchronous\xspace}
\newcommand{\sync}{synchronous\xspace}
\newcommand{\Sync}{Synchronous\xspace}

\newcommand{\ev}[1]{e_\textit{#1}}
\newcommand{\da}[1]{d_\textit{#1}}
\newcommand{\tsys}{t}
\newcommand{\eval}{\mathsf{det}}
\newcommand{\evalt}{\mathsf{det}_T}
\newcommand{\trans}{\leadsto}
\newcommand{\transt}{\overset{\scriptscriptstyle T}{\rule{0pt}{.6ex}\smash{\leadsto}}}
\newcommand{\act}{\mathcal{CS}_0}
\newcommand{\uml}[1]{\textsf{\small #1}}

\newcommand{\mysubsubsection}[1]{\medskip\noindent\textbf{#1.}}

\usepackage[shortcuts]{extdash}
\def\figscaleoracles{0.88}

\journal{Journal}
\begin{document}

\begin{frontmatter}

\title{Which Event Happened First? \\ Deferred Choice on Blockchain Using Oracles\tnoteref{t1}}
\tnotetext[t1]{This research did not receive any specific grant from funding agencies in the public, commercial, or not-for-profit sectors.}

\author{Jan Ladleif\corref{cor1}}
\ead{jan.ladleif@hpi.uni-potsdam.de}
\author{Mathias Weske}
\ead{mathias.weske@hpi.uni-potsdam.de}
\address{Hasso Plattner Institute, University of Potsdam, Potsdam, Germany}

\cortext[cor1]{Corresponding author}

\begin{abstract}
First come, first served:
Critical choices between alternative actions are often made based on events external to an organization, and reacting promptly to their occurrence can be a major advantage over the competition.
In \gls{bpm}, such deferred choices can be expressed in process models, and they are an important aspect of process engines.
Blockchain-based process execution approaches are no exception to this, but are severely limited by the inherent properties of the platform:
The closed-world environment prevents direct access to external entities and data, and the passive runtime based entirely on atomic transactions impedes continual monitoring and detection of events.
In this paper we provide an in-depth examination of the semantics of deferred choice, and transfer them to environments such as the blockchain.
We introduce and compare several oracle architectures able to satisfy certain requirements, and show that they can be implemented using state-of-the-art blockchain technology.
\end{abstract}


\end{frontmatter}

\glsresetall

\section{Introduction}
\label{sec:introduction}

Noticing a low-level system malfunction, receiving an order from a customer, or a stock finally reaching the strike price---events are pervasive on every layer of abstraction in \gls{bpm}~\cite{weske2019business}.
Often being a part of complex structural and behavioral patterns, individual events directly govern how businesses and their core processes operate.
Reacting to events promptly and correctly has a significant impact on the business performance, which is further exacerbated when events occur external to the business and are publicly visible to all competitors.

In today's digital and automated businesses, it is the job of the process engine to keep track of such events and inform running process instances of event detections.
Traditionally being deployed on premise or in cloud systems, a process engine monitors external event sources continuously using the wide range of techniques at its disposal, for example via querying public interfaces or subscribing to news channels.
This ensures a timely handling of events and execution of processes compliant with their specification, which not only drives the business value of a process but might also carry a legal relevance.

Recent approaches at implementing core aspects of process engines as smart contracts on blockchains, however, face severe issues trying to achieve the same level of support for events.
While the integrity, immutability, and non-repudation guarantees of blockchains provide clear benefits~\cite{mendling2017opportunities}---especially in collaborative settings~\cite{weber2016untrusted}---, they cause a peculiar execution environment which is isolated by design and inherently driven by transactions.
That is, process engines neither have the same techniques at their disposal to detect external events or access external data sources, nor are they running constantly outside of discrete transactions resulting in delayed or even missed event detections.
This has led to minimal or non-existing support for external events in state-of-the-art blockchain-based process engines~\cite{ladleif2020external}.

The goal of this paper is to resolve these issues, and provide a tight integration of external events on blockchain despite the idiosyncratic limitations of the platform.
We will specifically consider one of the major structural patterns in this context, \emph{deferred choice}, in which not only the detection of events but also their ordering influences a process~\cite{russell2006control}.
To achieve this, we will analyze the core semantics of external events and deferred choice, and then adapt them to cope with the transaction-driven execution environment of the blockchain.
We will provide concrete architectures using oracles to circumvent the isolation and access external event data, which we assess within a prototypical implementation.

The paper is structured as follows:
We first introduce preliminaries from the fields of \gls{bpm} and blockchain technology in Sect.~\ref{sec:preliminaries}.
After detailing our guiding research questions in Sect.~\ref{sec:motivation}, we start with the definition of a formal execution semantics in Sect.~\ref{sec:semantics}.
We propose generic oracle architectures to implement the semantics in Sect.~\ref{sec:oracles}, and provide a concrete prototype in Sect.~\ref{sec:prototype} which forms the basis of an evaluation in Sect.~\ref{sec:evaluation}.
After giving an overview of related work in Sect.~\ref{sec:related}, we discuss our results and conclude in Sect.~\ref{sec:discussion}.

\section{Preliminaries}
\label{sec:preliminaries}

In this section, preliminary knowledge about \gls{bpm} and blockchain technology will be introduced, with a focus on events and how the limitations of blockchain affect them.

\subsection{Business Process Management and Events}
\label{subsec:bpm-and-events}
Businesses are driven by recurring processes, in which activities are performed to reach a certain business goal.
\gls{bpm} is concerned with making these processes palpable and support them during their entire lifecycle:
from initial identification to structured modeling and execution using a process engine and subsequent optimization~\cite{weske2019business}.

Events are an essential aspect of business processes and refer to ``points in time''~\cite[p.~85]{weske2019business} at which something happens, in contrast to activities with a particular duration.
Events can take many shapes:
The de facto industry standard for business process models, \gls{bpmn}, for example, includes more than 10 types of events~\cite{omg2013bpmn}.
From error events, which represent system errors interrupting the regular process execution, to compensation events, which start attempts at rolling back certain activities---there are many options to represent critical business scenarios.

During the execution of a process, a process engine has to detect events as they occur to correctly apply their effects in a timely fashion.
This is especially complex for events occurring \emph{external} to the process engine.
Common examples are events caused by explicit external actions, like receiving an order or inquiries from customers within messages.
External events can also be caused more implicitly based on properties of the larger execution environment of the process, such as detecting a condition becoming satisfied, e.g., a stock reaching a specific price~\cite{russell2006control}.
Timer events also fall into this category and rely on the current time of the environment~\cite{eder1999time}, allowing absolute or relative temporal constraints to be specified~\cite{cheikhrouhou2015temporal}.

External events are the core building blocks of one of the fundamental patterns in workflow modeling and \gls{bpm}: \emph{deferred choice}~\cite{russell2006control}.
Deferred choice describes situations in which the exclusive choice between alternative execution branches of a process depends on the operating environment, i.e., which of a set of external events is detected first.
As these events are caused by actions or circumstances outside the influence of the process engine, the choice is deferred until the first event is detected---essentially modeling a ``race condition where the first [e]vent that is triggered wins''~\cite[p.~298]{omg2013bpmn}.

\subsection{Blockchain and BPM}
\label{subsec:blockchain}
Advancing from its roots in cryptocurrencies~\cite{nakamoto2008bitcoin}, blockchain has since emerged as a dedicated software platform which especially lends itself to business processes~\cite{weber2016untrusted}.

\mysubsubsection{Blockchain Technology}
Blockchains are distributed ledgers which store and process data and transactions.
Smart contracts are programs whose code and state is persisted as data on the blockchain, and which can be called using transactions targeting their exposed functions\cite{xu2017taxonomy}.
The result of these transactions comes into effect during mining, a process in which transactions are ordered, executed, and bundled into a new block cryptographically linked to its predecessor.

Smart contracts and interactions with them enjoy all the benefits of blockchain technology:
Their integrity can be easily validated, the origin of transactions can not be repudiated, and the smart contract state can not be forged~\cite{xu2017taxonomy}.
This makes them a valuable system component in low trust but high stakes scenarios involving several organizations, such as business process collaborations.

As a consequence, however, smart contracts need to adhere to several restrictions caused by the blockchain's idiosyncratic properties.
For one, they operate in an isolated closed-world environment for integrity and traceability reasons, meaning they can not access any data or service outside the blockchain itself.
Secondly, smart contracts are inherently passive and are only executed within discrete transactions during the mining process.
This lends them a kind of transaction-driven character, as they will always be ``paused'' between transactions and can not implement ongoing behavior such as busy-waiting.

\begin{figure}
\centering
\subfloat[\Storage oracle]{
\includegraphics[scale=\figscaleoracles]{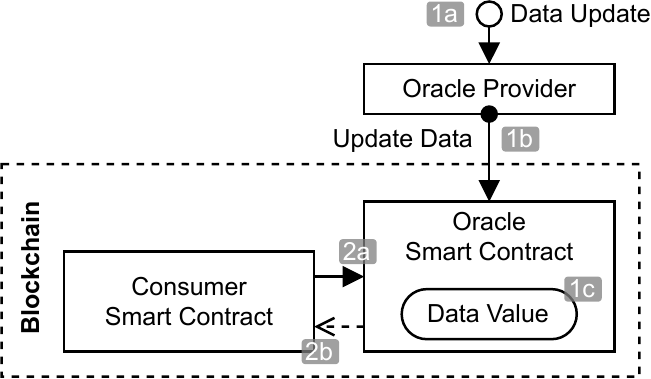}
\label{subfig:storage-oracle}}
\hfil
\subfloat[\Request oracle]{
\includegraphics[scale=\figscaleoracles]{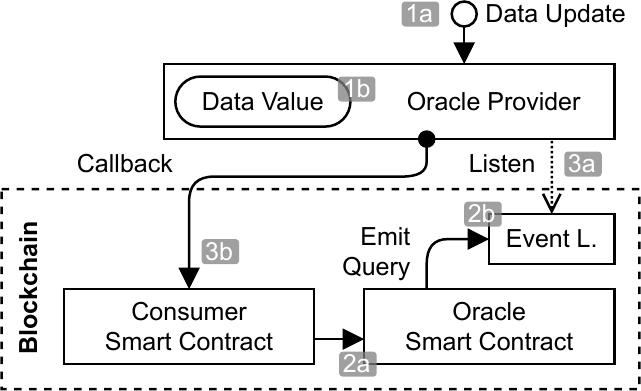}
\label{subfig:request-oracle}}
\caption{Architecture and behavior of the \storage and \request oracles}
\label{fig:storage-request-oracles}
\end{figure}

\mysubsubsection{Oracles}
In practice, \emph{oracles} are used to somewhat escape the isolation of blockchains and connect smart contracts to external entities~\cite{xu2018pattern}.
They generally consist of two components:
an off-chain oracle provider, which is not subject to the blockchain's restrictions and can freely access data, services, and entities;
and a publicly known oracle smart contract, which serves as a link between the consumers and the oracle provider.
Existing oracles usually use one of two basic architectures~\cite{albreiki2020trustworthy}:

The \emph{\storage oracle} (see Fig.~\ref{subfig:storage-oracle}) provides \sync data access by storing current data on the blockchain.
On each data update (1a) a transaction is sent (1b) to update a storage variable (1c) in the oracle smart contract.
Consumers can then query the oracle smart contract (2a), which directly responds with the last known data (2b).

The \emph{\request oracle} (see Fig.~\ref{subfig:request-oracle}) provides \async access to external data, avoiding the use of on-chain storage.
Instead, the oracle provider keeps track of the current data, for example by getting updates (1a) and storing them locally (1b).
Consumer smart contracts can call the oracle smart contract (2a), which emits the query in an event using the blockchain's event layer (2b).
The oracle provider picks up this event (3a), and sends the data to the consumer smart contract in a new transaction (3b).

\mysubsubsection{Blockchain-Based Process Engines}
The potential synergy of blockchain technology and \gls{bpm} was acknowledged early on~\cite{mendling2017opportunities}, shortly after first work was conducted to use smart contracts for secure and traceable process execution~\cite{weber2016untrusted}.
This has spawned an impressive amount of research, especially regarding collaborative processes which benefit considerably from the blockchain's security properties~\cite{garcia2020blockchain}.

In most approaches, one or more smart contracts keep track of the state of a process instance.
Actions within the process are funneled through the smart contract using transactions, which can directly advance the state of the process instance accordingly or be rejected if they do not conform to the process specification.
In any case, the blockchain provides a tamper-proof audit log~\cite{weber2016untrusted}.
Such process smart contracts need to implement the correct semantics of the original process model specification.
Whether this is achieved using code generation~\cite{weber2016untrusted,lopez2019caterpillar,ladleif2019modeling} or interpretation~\cite{lopez2019interpreted} is secondary, as long as each source process concept is mapped to an equivalent smart contract concept.

In practice, this mapping is a particular problem for external events and the deferred choice pattern---which we will elaborate on in the remainder of this paper.

\section{Deferred Choice on Blockchain}
\label{sec:motivation}

The deferred choice pattern relies on capabilities which the blockchain innately lacks---a continuous monitoring of the external environment, conflicting with the passive and isolated environment encapsulating smart contracts.
In this section, we will further illustrate these issues, and describe our approach in solving them.

\begin{figure}
  \centering
  \includegraphics[width=.85\linewidth]{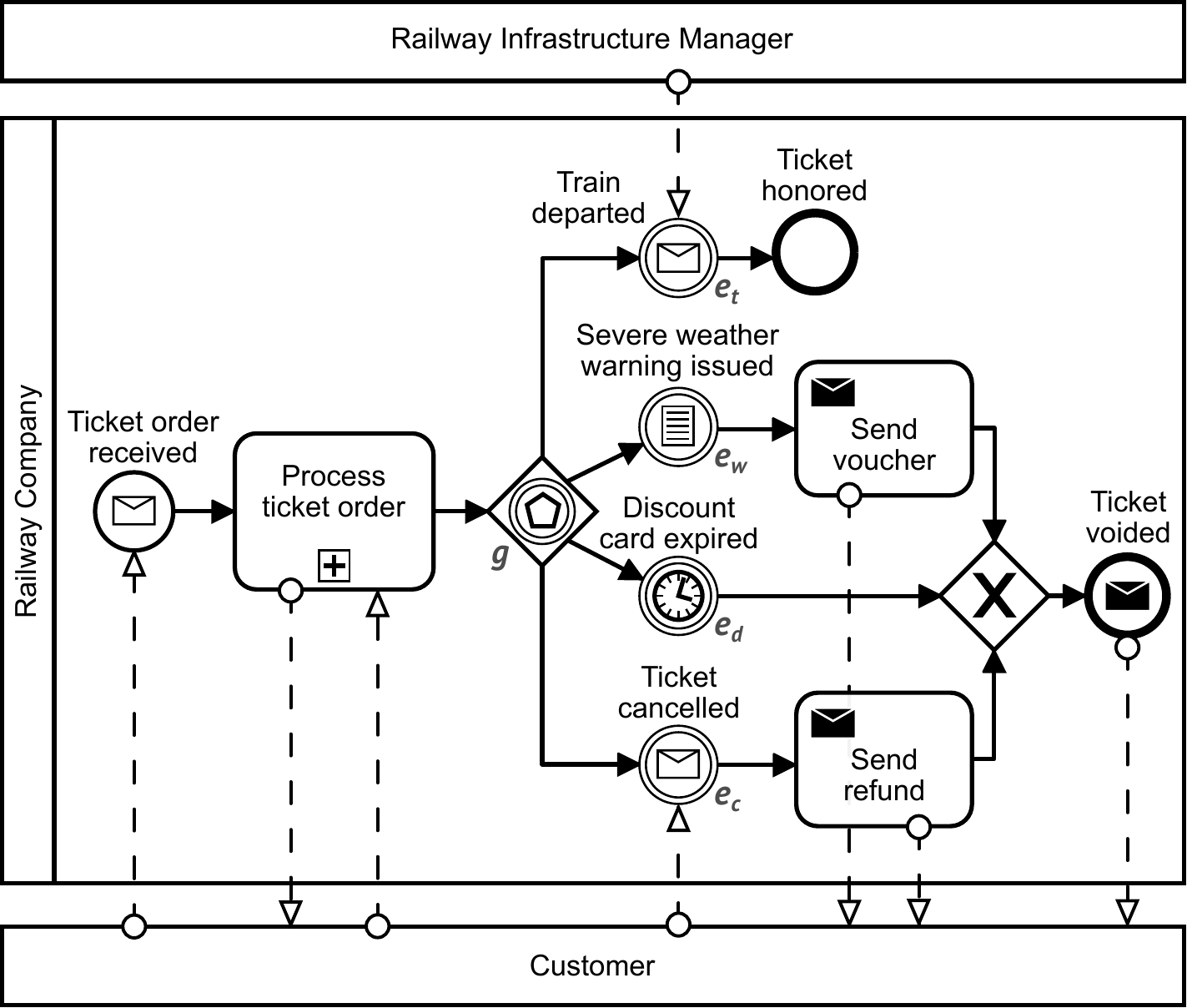}
  \caption{Running example of a deferred choice scenario}
  \label{fig:running-example}
\end{figure}

\subsection{Motivating Example}
\label{subsec:motivating-example}
Railway companies often sell tickets to their trains in advance, especially on long-haul routes.
Many companies also provide membership-based discount cards, which give customers a percentage off their ticket price.
Optimally, the customer books a ticket with their discount, arrives at the train station before departure, and the ticket is honored.
Figure~\ref{fig:running-example} shows this scenario modeled as a \gls{bpmn} collaboration diagram between the railway company and external actors, in which the train's departure is expressed by a message ($\ev{t}$) from the \gls{rim}.

Of course, reality is often more complex, and the diagram contains several exception paths after the event-based gateway $g$---the canonical representation of a deferred choice in \gls{bpmn}~\cite{omg2013bpmn}.
The customer may cancel their ticket ($\ev{c}$), a severe weather might prohibit the train's departure ($\ev{w}$), or the discount card may expire before the train departs ($\ev{d}$).
In all cases, the ticket is voided with a variable display of generosity from the railway company in making amends for the loss of the ticket.

\subsection{Detecting External Events on Blockchain}
\label{subsec:problem-statement}

In blockchain-based process execution, the business logic expressed in the model in Fig.~\ref{fig:running-example} is encapsulated in a smart contract which assumes the role of the process engine.
Within some transaction to this smart contract, the process execution will reach the gateway $g$ and the deferred choice is started.
The transaction finishes, and the smart contract lies dormant.
In Fig.~\ref{fig:problem-timeline}, we call this transaction $tx$.
Additionally, the figure shows how the external environment might develop while the smart contract is inactive:

\begin{figure}
  \centering
  \includegraphics[width=.65\linewidth]{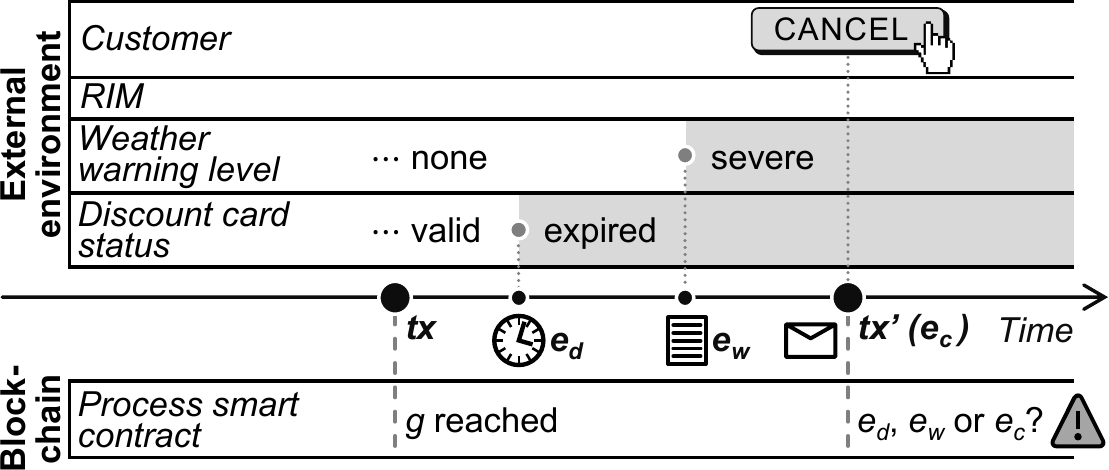}
  \caption{Indeterminate behavior of a process smart contract due to temporal gaps between transactions}
  \label{fig:problem-timeline}
\end{figure}

First, the customer's discount card expires.
Then, the meteorological service issues a severe weather warning.
And finally, the customer---perhaps in an attempt to get a refund on the soon-to-be voided ticket---sends a message asking for a ticket cancellation.
Being an explicit action by a process participant, this message is issued via a transaction $tx'$, finally waking the smart contract up again.
In $tx'$, the smart contract is able to detect $\ev{d}$, since the deadline has passed; it acquires the current weather warning level using an oracle and is able to detect $\ev{w}$; and the transaction itself leads to the detection of $\ev{c}$.
However, it is impossible for the smart contract to decide which event came first and should win the race---which is acutely relevant, since a refund may erroneously be sent.

This example illustrates that the deferred choice pattern is challenging to implement for smart contracts for two reasons:
First, the passiveness of smart contracts makes it difficult to perform continuous tasks which process engines have to perform, like monitoring implicit events---$\ev{d}$ and $\ev{w}$ are missed because there is by default no system automatically issuing a transaction.
Second, existing oracle architectures only provide access to current states of external data sources, making it impossible to determine when exactly a condition became fulfilled.

\subsection{Research Questions}
\label{subsec:research-questions}
The goal of this paper is to enable the correct execution of deferred choice patterns within smart contracts on blockchain to enable full compliance in blockchain-based process execution frameworks.
To this end, we plan to address the issues we have identified by answering the following research questions:

\begin{enumerate}[RQ 1.]
  \item How can the intended semantics of deferred choice be achieved in a transaction-driven environment like the blockchain?
  \item How can the resulting semantics be implemented with the help of practical oracle architectures?
\end{enumerate}

The first questions will be the focus of Sect.~\ref{sec:semantics}, while the second question will be tackled in Sect.~\ref{sec:oracles} and onward.

\section{Deferred Choice Execution Semantics}
\label{sec:semantics}

To gauge the impact of transaction-driven execution platforms like blockchains, we introduce a formal semantics of deferred choice from a continual perspective.
It comprises of two layers as suggested in literature~\cite{russell2006control}:
the operating environment in which events explicitly occur or whose properties make events implicitly occur,
and the deferred choice detecting those events.
We then adapt the semantics to the notion of transactions.

\subsection{Operating Environment}
A deferred choice represents a local decision based on external factors outside the influence of the process engine (see Sect.~\ref{subsec:bpm-and-events}), which are summarized as the environment:

\begin{definition}[environment]
The (operating) environment encapsulates all actors, systems, and information external to the process engine which are necessary to execute a deferred choice.
In particular, this includes a set $D$ of external variables.
\end{definition}

Clearly, the environment is tailored to an individual application of the deferred choice pattern.
For instance, the environment of the deferred choice $g$ in the train ticket process (see Fig.~\ref{fig:running-example}) comprises of two external actors: the customer and the \gls{rim}.
There also has to be an external variable, say $D=\{\da{w}\}$, holding the current weather warning level, which would be stored within a database at the meteorological service.
Lastly, the current time is logically part of the environment, and is used to keep track of the discount card expiration.

To keep our definitions concise, we assume in the following that we consider exactly one arbitrary but fixed deferred choice and its associated environment.
Likewise, all examples will refer to the deferred choice $g$ from our motivating example (see Sect.~\ref{subsec:motivating-example}).

While explicit events are usually actively triggered by external actors, implicit events rely on certain environment properties at a specific point in time.
For example, conditional events are bound to the values of an external variable, and timer events to the current environment time, all subsumed in the environment state:

\begin{definition}[environment state]
Let $\mathbb{D}$ be the domain of all possible values of external variables.
Then an environment state $s=(\tsys, \nu)$ is a tuple consisting of a timestamp $\tsys\in\mathbb{N}$ and a valuation function $\nu: D\rightarrow \mathbb{D}$ assigning a value to each variable.
The set of all environment states is called $\mathcal{ES}$.
\end{definition}

We use the natural numbers $\mathbb{N}$ as timestamps, which is a common abstraction found in information systems, and results in a discrete and uniform time domain.
Naturally, the state of the environment changes and evolves as time passes:

\begin{definition}[successors]
\label{def:successors}
Let $s=(t,\nu), s'=(t',\nu') \in \mathcal{ES}$ be environment states.
Then $s'$ is a successor of $s$, or $s\rightarrow s'$, iff $\tsys' = \tsys + 1$.
\end{definition}

\begin{table}
\centering
\caption{Operating environment, event detections and example deferred choice traces for the train ticket scenario}
\label{tab:example-timelines}
\newcommand{\wrap}[1]{\hskip15pt\makebox[0pt]{#1}\hskip15pt}
\wrap{
\small
\begin{tabular}{l||c@{}c*6{cc}c}

\hline\rowcolor{black!10} \multicolumn{16}{l}{\textbf{Environment trace}} \\\hline Environment states $s_i$& $ \cdots $& \wrap{$\rightarrow$}& \wrap{$s_1$}& \wrap{$\rightarrow$}& \wrap{$s_2$}& \wrap{$\rightarrow$}& \wrap{$s_3$}& \wrap{$\rightarrow$}& \wrap{$s_4$}& \wrap{$\rightarrow$}& \wrap{$s_5$}& \wrap{$\rightarrow$}& \wrap{$s_6$}& \wrap{$\rightarrow$}& $ \cdots $ \\\hline Timestamp $\tsys_i$& $ \cdots $& & \wrap{73}& & \wrap{74}& & \wrap{75}& & \wrap{76}& & \wrap{77}& & \wrap{78}& & $ \cdots $ \\\hline Valuation $\nu_i(\da{w})$& $ \cdots $& & \wrap{0}& & \wrap{1}& & \wrap{1}& & \wrap{1}& & \wrap{2}& & \wrap{2}& & $ \cdots $ \\\hline\rowcolor{black!10} \multicolumn{16}{l}{\textbf{Event detections}} \\\hline $\eval(\ev{d}, s_1, s_i) ~\hat=~ 76 \leq \tsys_i$& $ \cdots $& & & & & & & & \wrap{true}& & \wrap{true}& & \wrap{true}& & $ \cdots $ \\\hline $\eval(\ev{w}, s_1, s_i) ~\hat=~ \nu_i(\da{w}) \geq 2$& $ \cdots $& & & & & & & & & & \wrap{true}& & \wrap{true}& & $ \cdots $ \\\hline $\eval(\ev{c}, s_1, s_i)$& $ \cdots $& & & & & & & & & & & & \wrap{true}& & $ \cdots $ \\\hline $\eval(\ev{t}, s_1, s_i)$& $ \cdots $& & & & & & & & & & & & & & $ \cdots $ \\\hline\rowcolor{black!10} \multicolumn{16}{l}{\textbf{Deferred choice traces}} \\\hline (a) Continual& $  $& \wrap{$\act\ni~~~~~~$}& \wrap{$(s_1, s_1, nil)$}& \wrap{$\trans$}& \wrap{$(s_1, s_2, nil)$}& \wrap{$\trans$}& \wrap{$(s_1, s_3, nil)$}& \wrap{$\trans$}& \wrap{$(s_1, s_4, \ev{d})$}& \wrap{$~~\in\mathcal{F}$}& & & & & $  $ \\\hline (b) Transaction-driven, history& $  $& \wrap{$\act\ni~~~~~~$}& \wrap{$(s_1, s_1, nil)$}& & & & & \wrap{$\transt$}& & & & & \wrap{$(s_1, s_6, \ev{d})$}& \wrap{$~~\in\mathcal{F}$}& $  $ \\\hline (c) Transaction-driven, pub/sub& $  $& \wrap{$\act\ni~~~~~~$}& \wrap{$(s_1, s_1, nil)$}& \wrap{$\transt$}& \wrap{$(s_1,s_2, nil)$}& & & \wrap{$\transt$}& & & \wrap{$(s_1, s_5, \ev{d})$}& \wrap{$~~\in\mathcal{F}$}& & & $  $ \\\hline

\end{tabular}
}
\end{table}

An example trace of such environment states is shown in the upper part of Tab.~\ref{tab:example-timelines}, e.g., $s_1=(\tsys_1, \nu_1)$ with $\tsys_1=73$ and $\nu_1(\da{w})=0$.
The timestamps grow by 1 in each successive state, and the valuation function changes unpredictably.

\subsection{Deferred Choice}
The structure of a deferred choice is simple:

\begin{definition}[deferred choice]
A deferred choice is made between a non-empty set $E$ of external events.
\end{definition}

These events can be both explicit and implicit.
In the train ticket example, for instance, the deferred choice $g$ is given by $E=\{ \ev{t}, \ev{w}, \ev{d}, \ev{c} \}$.

It is helpful to think of a deferred choice as a ``race'' with a single winner.
Like for all races, there are two important milestones: when it starts and when it ends.
A deferred choice starts or is \emph{activated} when it is encountered during process execution, e.g., gateway $g$ is reached; and ends once a winning event is detected, e.g., the expiration of the discount card $\ev{d}$.
The detection of events is contingent on the current environment state as well as the one at activation:

\begin{definition}[event detection]
\label{event-detection}
Given a deferred choice $E$, $\eval: E \times \mathcal{ES} \times \mathcal{ES} \rightarrow \{ true, false \}$ is the Boolean detection function with $\eval(e, s_a, s)=true$, iff an event $e$ can be detected in environment state $s$ assuming activation happened in $s_a$.

In extension, $E_\eval: \mathcal{ES}\times\mathcal{ES}\rightarrow 2^E$ with $E_\eval(s_a, s) := \{ e \in E ~|~ \eval(e,s_a,s) \}$ is the set of all implicit events which can be detected under these circumstances.
\end{definition}

The concrete definition of $\eval$ depends on the type of event.
Several examples can be seen in the middle part of Tab.~\ref{tab:example-timelines}.
For the implicit events, expressions over the environment states can be given:
$\eval(\ev{d}, s_a, s) ~\hat=~ \tsys \geq 76$ means that the discount card expires at timestamp 76;
and $\eval(\ev{w}, s_a, s) ~\hat=~ \nu(\da{w}) \geq 2$ means that the severe weather warning event $\ev{w}$ is detected as the corresponding external variable $\da{w}$ reaches the value 2.
Thus, both these events can be detected in $s_5$ assuming activation at $s_1$: $E_\eval(s_1, s_5) = \{ \ev{d}, \ev{w} \}$.
Explicit events, on the other hand, are based on concrete actions in the environment, e.g., messages sent by external actors, and follow no discernible pattern.

\subsection{Continual Execution Semantics}
\label{subsec:continual-semantics}
In a race, a winner is determined immediately upon their reaching of the goal.
Achieving this behavior for deferred choice (see Sect.~\ref{subsec:bpm-and-events}) requires a continual observation of the environment to detect events as soon as possible.
To express this formally, we propose a state transition system based on the evolution of the deferred choice alongside the environment state:

\begin{definition}[deferred choice state]
Given a deferred choice $E$, a deferred choice state $(s_a, s, e) \in \mathcal{ES} \times \mathcal{ES} \times (E \cup \{nil\})$ is a tuple with $s_a$ the environment state at activation, $s$ the last observed environment state, and $e$ the winning event or $nil$, if no event has won yet.
The set of all such states is called $\mathcal{CS}$.
\end{definition}

The initial states of the transition system are determined by the environment state when the deferred choice is activated and the race starts:

\begin{definition}[initial states]
The set $\act\subseteq\mathcal{CS}$ of all initial states of a deferred choice $E$ is given by
\begin{align*}
  \act := \{ (s,s,e) ~|~ s\in\mathcal{ES} \wedge \big( & e\in E_\eval(s, s) ~\vee \\
  & (E_\eval(s, s) = \emptyset \wedge e=nil) \big) \}
\end{align*}
\end{definition}

Notably, an initial state might already be the end of the race:
If events can be detected immediately, one of them must be chosen.
There is no sense of priority; for example (see Tab.~\ref{tab:example-timelines}), $(s_5, s_5, \ev{d})$ and $(s_5, s_5, \ev{w})$ are the valid initial states when starting in $s_5$, but $(s_1, s_1, nil)$ is the only one when starting in $s_1$.
The race ends when a winner is found:

\begin{definition}[final states]
$\mathcal{F} := \{ (s_a, s, e)\in\mathcal{CS} ~|~ e\neq nil \}$ is the set of all final deferred choice states, i.e., those in which an event has won.
\end{definition}

Using these notions, we can finally define the operational semantics of deferred choice using an unlabeled terminal transition system with initial states~\cite{plotkin1981sos}:

\begin{definition}[continual transition system]
\label{def:continual-transition-system}
Given a deferred choice $E$, the transition system $\left\langle \mathcal{CS}, \trans, \act, \mathcal{F} \right\rangle$ with a transition relation $\trans\,\subseteq\mathcal{CS}\times\mathcal{CS}$ such that
\[
  \begin{array}{cl>{\scriptstyle}r}
    \multicolumn{3}{l}{(s_a,s,nil) \trans (s_a,s',e)} \\
    \Longleftrightarrow &
    s \rightarrow s' ~\wedge & \text{(i)} \\
    & (e \in E \implies e\in E_\eval(s_a, s')) ~\wedge & \text{(ii)} \\
    & (e = nil \implies E_\eval(s_a, s') = \emptyset) & \text{(iii)} \\
  \end{array}
\]
describes its continual execution semantics.
\end{definition}

The transition relation $\trans$ includes several constraints.
For one, (i) the two referenced environment states need to be direct successors.
There can be no gap in time between them, expressing the continual nature of transitions.
Further, (ii) if an event wins, it must be among those detected in the new environment state $s'$.
Lastly, (iii) the winner may only remain undecided if no event could have been detected.

The bottom part of Tab.~\ref{tab:example-timelines} shows an example trace (a) of the continual semantics, starting with the state $(s_1, s_1, nil)\in\act$.
The state evolves alongside the environment, until $\ev{d}$ is detected and the transition system arrives in the final state $(s_1, s_4, \ev{d})\in\mathcal{F}$.

\subsection{Transaction-Driven Execution Semantics}
The continual execution semantics works under the assumption that each environment state is observed:
Deadlines being reached or conditions becoming satisfied will immediately be registered, and corresponding events be detected.

However, this is not the case in a transaction-driven environment like the blockchain.
Each change of the deferred choice state, that is activating and transitioning to new states with $\trans$, needs to be contained within a blockchain transaction.
Time passes between transactions, in which the smart contract storing the state lies dormant.
This directly leads to environment states being missed, violating rule (i) of the continual transition relation $\trans$.

Thus, $\trans$ needs to be adapted.
Rule (i) needs to be relaxed to allow for gaps, and the other rules modified to cope with those gaps.
In a first step, we reconsider the event detection function $\eval$.
Instead of just checking \emph{whether} an implicit event may be detected in a single environment state, we extend it to return the time at which an event was first detected---that is, when it crossed the finish line in the race:

\begin{definition}[timed event detection]
\label{def:timed-event-detection}
Let $\top\in\mathbb{N}$ be a fixed and sufficiently large timestamp arbitrarily far in the future that will realistically never be reached in the environment.

Given a deferred choice $E$, the function $\evalt: E \times \mathcal{ES} \times \mathcal{ES} \rightarrow \mathbb{N}$ determines for $\evalt(e,s_a,s)$ the \emph{earliest} timestamp at which $e$ could have been detected starting from the environment state $s_a$ up to $s$.
If no such timestamp exists, $\top$ is returned as an indicator that the event could not have been detected.

In extension, $t_T:\mathcal{ES}\times\mathcal{ES}\rightarrow\mathbb{N}$ with $t_T(s_a, s) := \min\{ \evalt(e, s_a, s) ~|~ e \in E \}$ is the earliest detection time of any event, or $\top$ if none was detected at all.
\end{definition}

For example, in the train ticket scenario (see Tab.~\ref{tab:example-timelines}), the discount card first expired at timestamp 76, leading to, e.g., $\evalt(\ev{d}, s_1, s_6)=76$.
It is also the first event to occur, thus $t_T(s_1,s_6)=76$ holds as well.
Assuming that such a function exists---which we will qualify in the next section---, we can devise of a new transition system:

\begin{definition}[transaction-driven transition system]
\label{def:transaction-driven-transition-system}
Given a deferred choice $E$, the transition system $\left\langle \mathcal{CS}, \transt, \act, \mathcal{F} \right\rangle$ with a transition relation $\transt\,\subseteq\mathcal{CS}\times\mathcal{CS}$ such that
\[
  \begin{array}{cl>{\scriptstyle}r}
    \multicolumn{3}{l}{(s_a,s,nil) \transt (s_a,s',e)} \\
    \Longleftrightarrow &
    t < t' ~\wedge & \text{(i)} \\
    & (e\in E \implies \evalt(e, s_a, s') = t_T(s_a, s') \neq \top) & \text{(ii)} \\
    & (e = nil \implies t_T(s_a, s') = \top) & \text{(iii)}
  \end{array}
\]
describes its transaction-driven execution semantics.
\end{definition}

The transaction-driven transition relation $\transt$ only requires that $s'$ is after $s$ and not a direct successor anymore in rule (i)---which allows for the mentioned gaps.
Rule (ii) expresses that if an event is chosen as the winner, it must have been detected and no event may have been detected earlier.
Rule (iii) checks that the winner remains undecided only if no event could have been detected at all.

Speaking in terms of the race analogy, the transaction-driven semantics basically corresponds to the referee determining the time each participant has finished after the race, and retroactively deciding on the winner.
Table~\ref{tab:example-timelines} shows example traces (b) and (c) for this behavior.
In trace (b), we directly transition from the initial state $(s_1, s_1, nil)$ to $(s_1, s_6, \ev{d})$, correctly identifying $\ev{d}$ as the winner.
Trace (c) arrives at the same conclusion.

\subsection{Timed Event Detection on Blockchain}
\label{subsec:timed-detection}
While the transaction-driven transition relation $\transt$ ostensibly solves the issues of deferred choice on blockchain, it relies on a major assumption: that a timed detection function $\evalt$ exists.
The actual definition of such a function is needed for an implementation, however, and may be non-obvious depending on the type of event that is concerned.
In the scope of this paper, we consider three types of events:

\mysubsubsection{Message Events}
In blockchain-based process execution, message events usually get translated to transactions as shown in Fig.~\ref{fig:problem-timeline}.
The detection of a message event is then equivalent to the corresponding transaction being mined, which can easily be retrieved by the smart contract.
$\evalt$ for message events thus comes down to determining whether the current transaction directly corresponds to a message.

\mysubsubsection{Timer Events}
Since the exact deadline for absolute and delay for relative timers is known, it is easy to retroactively determine when they should have been detected first.
Given, for example, a relative timer event $e$ with a delay $\delta$ and a detection function of the form $\eval(e, s_a, s) ~\hat=~ \tsys_a + \delta \leq \tsys$, we can directly derive $\evalt$:
\[
  \evalt(e, s_a, s) \,\hat=
  \left\lbrace\begin{array}{ll}
    \tsys_a + \delta & \text{if } \tsys_a + \delta \leq \tsys \\
    \top & \text{otherwise}
  \end{array}\right.
\]
If the delay has passed since activation, it has done so exactly $\tsys_a+\delta$.
For absolute timer events, $\evalt$ looks similar.

\mysubsubsection{Conditional Events}
The valuation function $\nu$ may change arbitrarily between environment states.
There is no way to deduce the intermediate values of external variables from the environment states at activation and transitioning alone.
Thus, $\evalt$ can not be specified for conditional events without additional assumptions.
We suggest two approaches to this end:

The \emph{\history approach} assumes the intermediate environment states $s_a=s_1\rightarrow s_2\rightarrow\cdots\rightarrow s_n=s$ and thus all valuations of the external variables are available.
Using a simple search, one can then find the earliest timestamp at which the continual detection function $\eval$ returned $true$:
\[
  \evalt(e, s_a, s) \hat=
  \left\lbrace\begin{array}{l}
    \top, \text{ if } \nexists i: \eval(e, s_a, s_i) \\
    \min\{ t_i ~|~ \eval(e, s_a, s_i) \}, \text{ otherwise}
  \end{array}\right.
\]

The \emph{\publish approach} moves responsibility towards the environment:
We consider a system in which the deferred choice smart contract subscribes to an external variable, and is notified of any changes with transactions.
In these transactions, the smart contract could be sure that they have received a complete picture of all conditional event occurences up to that point, and $\evalt(e, s_a, s_i)$ would be equal to $\eval(e, s_a, s_i)$ until an event detection.

A core difference between the approaches is how quickly conditional events are detected after the occur.
For the \history approach, this entirely depends on the timing of transactions, as events can be detected retroactively.
The \publish approach, conversely, ensures a very timely reaction to conditional events since we assume an active notification.
In contrast to the other event types, though, both the approaches for conditional events currently lack support in practice, which we will examine in the next section.

\section{Extended Oracle Architectures}
\label{sec:oracles}

Oracles are needed to implement conditional events within smart contracts, since they rely on the values of external variables.
In addition, these oracles must provide certain functionality to support the transaction-driven semantics---that is, they must either deliver historical data, or provide \publish services (see Sect.~\ref{subsec:timed-detection}).
However, existing oracle architectures like \storage and \request are not capable of either.
We propose a set of purposeful extensions and refinements of existing architectures~\cite{ladleif2020external,xu2018pattern} precisely targeting these capabilities.

\begin{figure}
\centering
\subfloat[\Sync on-chain \history oracle]{
\includegraphics[scale=\figscaleoracles]{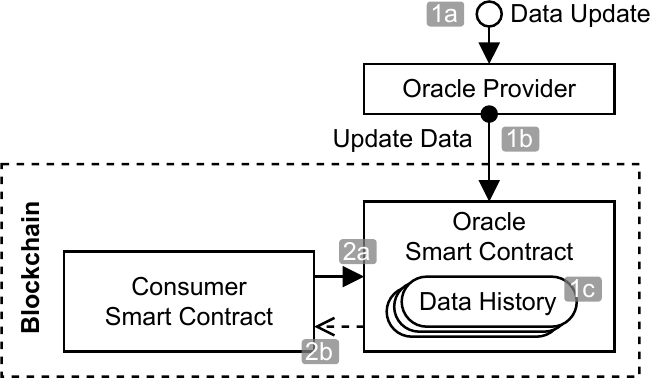}
\label{subfig:on-history-oracle}}
\hfil
\subfloat[\Async off-chain \history oracle]{
\includegraphics[scale=\figscaleoracles]{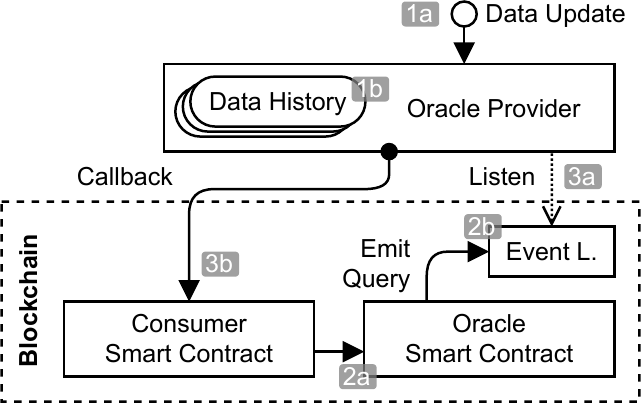}
\label{subfig:off-history-oracle}}
\caption{Architecture and behavior of the \history oracles}
\label{fig:history-oracles}
\end{figure}

\subsection{History Oracles}
\History oracles allow consumers to obtain not only the present value of an external data variable, but also a history of past values.
This makes it possible to implement the history approach at timed event detection explained in Sect.~\ref{subsec:timed-detection}.
We introduce two variants of the \history oracle:

\mysubsubsection{Architecture}
The \sync on-chain \history oracle (see Fig.~\ref{subfig:on-history-oracle}) is a direct extension of the existing \storage oracle (see Fig.~\ref{subfig:storage-oracle}).
The oracle provider observes the external data variable, and on data updates (1a) sends a transaction containing the new value to the oracle smart contract (1b).
Instead of overwriting the old value, however, the new value is added to an on-chain database holding a full history of the data (1c).
A consumer can then query a slice of this historical data (2a) and immediately receive result (2b).

The \async off-chain \history oracle (see Fig.~\ref{subfig:off-history-oracle}) likewise extends the existing \request oracle (see Fig.~\ref{subfig:request-oracle}).
Again, the oracle provider listens to updates of the external variable (1a), but stores a full history of the valuation off-chain (1b).
A consumer smart contract may now submit a query for a slice of the data to the oracle smart contract (2a), which is emitted using the blockchain's event layer (2b).
On receiving the query (3a), the oracle provider subsequently provides the requested data in a new transaction (3b).

\begin{table}
\centering
\small
\caption{Interfaces of the oracles from the perspective of the consumer}
\label{tab:oracle-interfaces}
\begin{tabular}{l|l||l|l}\hline

\emph{\bfseries Variant} &
\emph{\bfseries Oracle pattern} &
\multicolumn{2}{l}{\emph{\bfseries Interface domains}} \\\cline{3-4}

&
&
\emph{Parameters} &
\emph{Result} \\\hline\hline

Regular &
\Storage, \req &
\emph{none} &
$\mathbb{D}$ \\\cline{2-4}

&
\History &
$\mathbb{N}$ &
$(\mathbb{N}\times\mathbb{D})^*$ \\\cline{2-4}

&
\Publish &
\emph{none} &
$\mathbb{D}$ \\\hline

Condi- &
\Storage, \req &
$\textsc{EXPR}$ &
$\{ true, false \}$ \\\cline{2-4}

tional &
\History &
$\mathbb{N}\times\textsc{EXPR}$ &
$\mathbb{N}$ \\\cline{2-4}

&
\Publish &
$\textsc{EXPR}$ &
\emph{none} \\\hline

\end{tabular}
\end{table}

\mysubsubsection{Interfaces}
Table~\ref{tab:oracle-interfaces} shows the interfaces of all oracle architectures described in this paper in terms of the formalization in Sect.~\ref{sec:semantics}.
Regular \storage and \request oracles, for instance, do not require any input from the consumer smart contract, and return the current value of the external variable from $D$ they target.

Regular \history oracles are supplied one parameter, a timestamp from $\mathbb{N}$ at which to start the slice of historical values.
As an output, consumers will get a list of timestamped data values.
Specific points in time can then be found, for example, using a simple binary search.

\subsection{Publish-Subscribe Oracles}
As the name suggests, \publish oracles employ the well-known \publish software pattern, and let consumers subscribe to a specific external variable to actively notify them immediately of any changes.
The effect is that no historical data needs to be stored, which corresponds to the \publish approach presented in Sect.~\ref{subsec:timed-detection}.
However, the onus of providing those updates in time is shifted to an off-chain component, in this case the oracle provider.

\begin{figure}
\centering
\includegraphics[scale=\figscaleoracles]{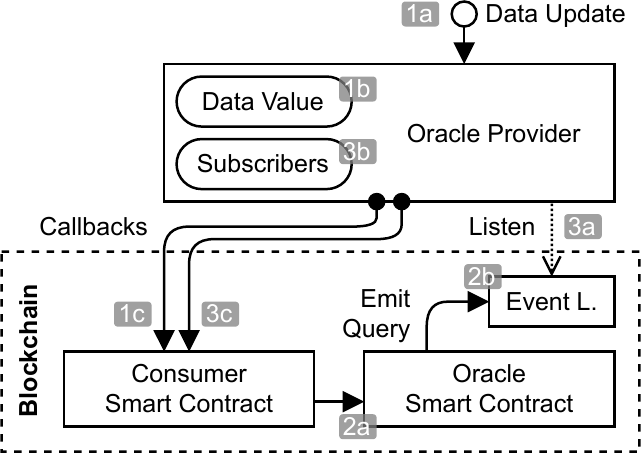}
\caption{Architecture and behavior of the \publish oracle}
\label{fig:publish-oracle}
\end{figure}

\mysubsubsection{Architecture}
Figure~\ref{fig:publish-oracle} shows the architecture of the \publish oracle.
The off-chain oracle provider observes the external data variable, and on updates (1a) stores the most recent value locally (1b) and also provides it to all subscriber smart contracts already registered (1c).
To subscribe, a consumer smart contract queries the oracle smart contract (2a) which emits an appropriate event (2b).
This event is picked up by the oracle provider (3a), which adds the consumer smart contract to the list of subscribers (3b) and also provides the current value immediately to avoid any gaps (3c).

\mysubsubsection{Interfaces}
From an interface perspective, the regular \publish oracle is equal to the traditional oracles (see Tab.~\ref{tab:oracle-interfaces}).
It does not require additional input, and provides a singular value each time an update occurs.
Note that we do not consider unsubscription in detail in the scope of this paper.
However, it could trivially be implemented by exposing an unsubscription function, which leads to the consumer smart contract being removed from the list of subscribers.

\subsection{Conditional Oracle Variants}
The oracle architectures above provide a strict separation of responsibilities when it comes to evaluating conditional events: the oracle provides the current value of the external variable, and the process smart contract evaluates specific conditions based on these values locally.
This pattern is realistic in that it protects the business knowledge of an organization---conditions themselves may be confidential and stored in protected areas of the blockchain.

However, the cost of operating a system on a blockchain platform heavily depends on both the amount of transactions being performed as well as the size of their payload, owing to the limited amount of storage space available~\cite{wood2014ethereum}.
Thus, we propose conditional oracles as a trade-off between confidentiality and cost by externalizing the evaluation of the condition attached to a conditional event to the oracle, either within its on-chain or off-chain infrastructure.
The goal is to cut down on the number of necessary transactions and the amount of data being exchanged.

The conditional variants alter the interfaces of the oracles as shown in the lower half of Tab.~\ref{tab:oracle-interfaces}.
As an additional input, an expression from the domain $\textsc{EXPR}$ of all possible expressions is given.
We do not specify the structure of those expressions in detail, but require them to yield a Boolean result and only reference external variables the specific oracle provides.

The output will then reflect the evaluation of the condition on the basis of the values of the external variable.
This is particularly interesting for \history oracles, which either return the earliest timestamp within the bounds given by the start timestamp at which the condition was fulfilled, or $\top$ to express the condition never evaluated to $true$---essentially reproducing the detection function given in Sect.~\ref{subsec:timed-detection}.
The conditional \publish oracle, on the other hand, returns no specific value at all.
Instead, a transaction is sent as a signal as soon as the condition becomes $true$.
Overall, this reduces both the amount of transactions needed as well as the payload size of the remaining transactions.

\section{Prototypical Implementation}
\label{sec:prototype}

To prove the feasibility of the transaction-driven semantics and oracle architectures, we developed a prototypical implementation.
It enables deployment of individual deferred choices containing message, timer, and conditional events on the blockchain, outside the context of their process.
The source code with all necessary information to reproduce our results is available online\footnote{\url{https://github.com/bptlab/blockchain-deferred-choice}}.

\begin{figure}
\centering
\includegraphics[width=\linewidth]{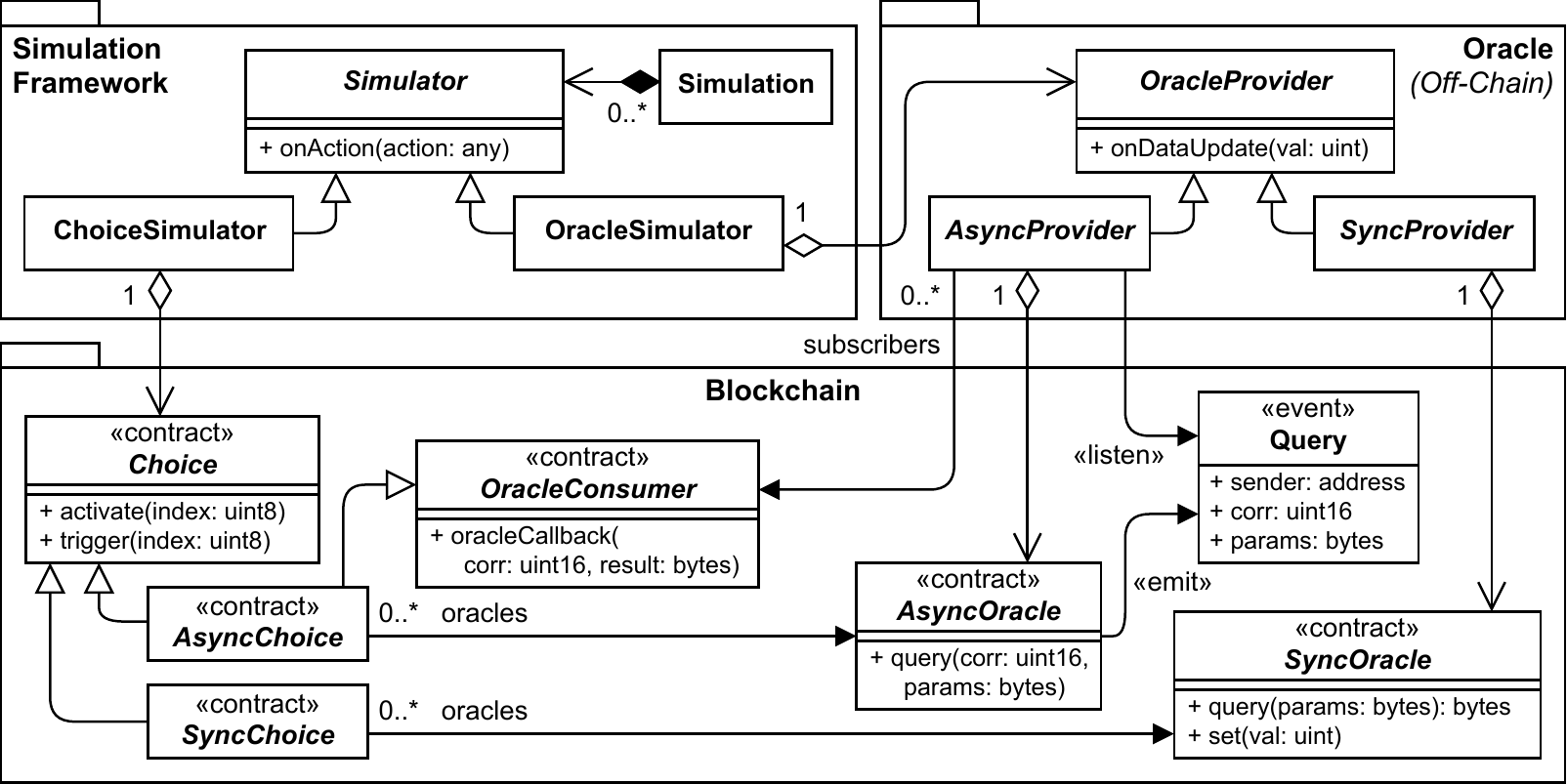}
\caption{Overall design of the prototype as a \gls{uml} class diagram}
\label{fig:prototype-architecture}
\end{figure}

\subsection{System Design}
The design of the prototype is narrowly aligned to that of the various oracle architectures.
There are three essential groups of components: (i) the blockchain smart contracts, (ii) the off-chain oracle providers, and (iii) an off-chain simulation framework.
Figure~\ref{fig:prototype-architecture} shows the overall structure as an \gls{uml} class diagram.
The non-standard \uml{contract} and \uml{event} stereotypes are used for smart contract classes and event types, respectively.

The general design goal was to achieve a level playing field for comparing the different semantics and oracles.
That is, we refrained from selective optimization of individual approaches.
At the same time, the system design is laid out to be as generic as possible, for example streamlining the interfaces to a minimum as explained later in this section.

A major theme of the prototype is a strict separation between \sync and \async oracles, which results in the dual class layout for the oracle providers, oracle smart contracts, and deferred choices.
For each concrete oracle architecture (see Sect.~\ref{sec:oracles}), e.g., the \sync on-chain \history oracle, there is a set of three corresponding concrete classes.
All custom code related to single or a group of oracle architectures can be moved to these classes.
In the following, we will walk through each part of the prototype.

\subsection{Blockchain Smart Contracts}
We use Ethereum~\cite{wood2014ethereum} as our target blockchain, since it provides all features necessary and its development ecosystem is well-maintained and accessible.
The core process logic is contained within smart contracts, which are implemented using the associated Solidity programming language.
For the data domain $\mathbb{D}$, we use \uml{uint256}, which is the largest static and atomic data type that Solidity allows for.
Timestamps are also stored using \uml{uint256}, and $\top$ is set to the type's largest value $2^{256}-1$ which is ``sufficiently'' (see Def.~\ref{def:timed-event-detection}) far---several hundred billion years---in the future.

\mysubsubsection{Oracles}
The smart contract classes \uml{AsyncOracle} and \uml{SyncOracle} contain the interfaces of the oracles (see Fig.~\ref{fig:prototype-architecture}).
Parameters and query results are encoded in raw byte arrays (\uml{bytes}) via the same mechanism Ethereum uses to encode transactions.
This allows us to use common interfaces for all oracles, from which data can be extracted according to specific interfaces (see Tab.~\ref{tab:oracle-interfaces}).
Larger payloads will incur a higher cost following the rules in the standard~\cite{wood2014ethereum}.

For \emph{\sync} oracles, the current value or historical values are stored on the blockchain and updated/appended via \uml{set}.
\Sync oracles may thus directly return a result upon \uml{query} being called.
For \emph{\async} oracles, the off-chain oracle provider stores data---no setter functions or on-chain storage are needed.
Instead, queries are emitted using a custom \uml{Query} event type containing all the required information for the off-chain oracle provider.
Consumers need to extend an additional \uml{OracleConsumer} contract to receive the later callback transaction.

A primary concern for \async oracles is achieving correlation:
A consumer needs to be able to link the transaction providing the query result to the query itself.
In practice, there are various strategies.
For example, Provable\footnote{\url{https://provable.xyz/}} returns a unique query ID, which is attached alongside the query result for later matching.
In our prototype, consumers choose an ID themselves in the form of the \uml{corr} value.

\mysubsubsection{Deferred Choice}
The contract class \uml{Choice} and its children implement the state and behavior of deferred choice.
Calling \uml{activate} via a transaction initially activates the deferred choice (equivalent to picking a valid initial state from $\act$), and \uml{trigger} is used to perform a step of the transition relation $\transt$ and potentially trigger an event and pick it as the winner.

In this context, the non-determinism of the transition system is an issue---if multiple events are detected at the same time, both are valid winners and there is no sense of priority (see Sect.~\ref{subsec:continual-semantics}).
Smart contracts are deterministic, though, and one event needs to be chosen.
To this end, we opted for a two-phase strategy:
Transactions may include a preferred event $e_i$ for each action which is chosen above all others if it is a valid winner.
Otherwise, the first valid winner is chosen in the order of the events' internal indices.

\begin{figure}
\centering
\includegraphics[scale=.68]{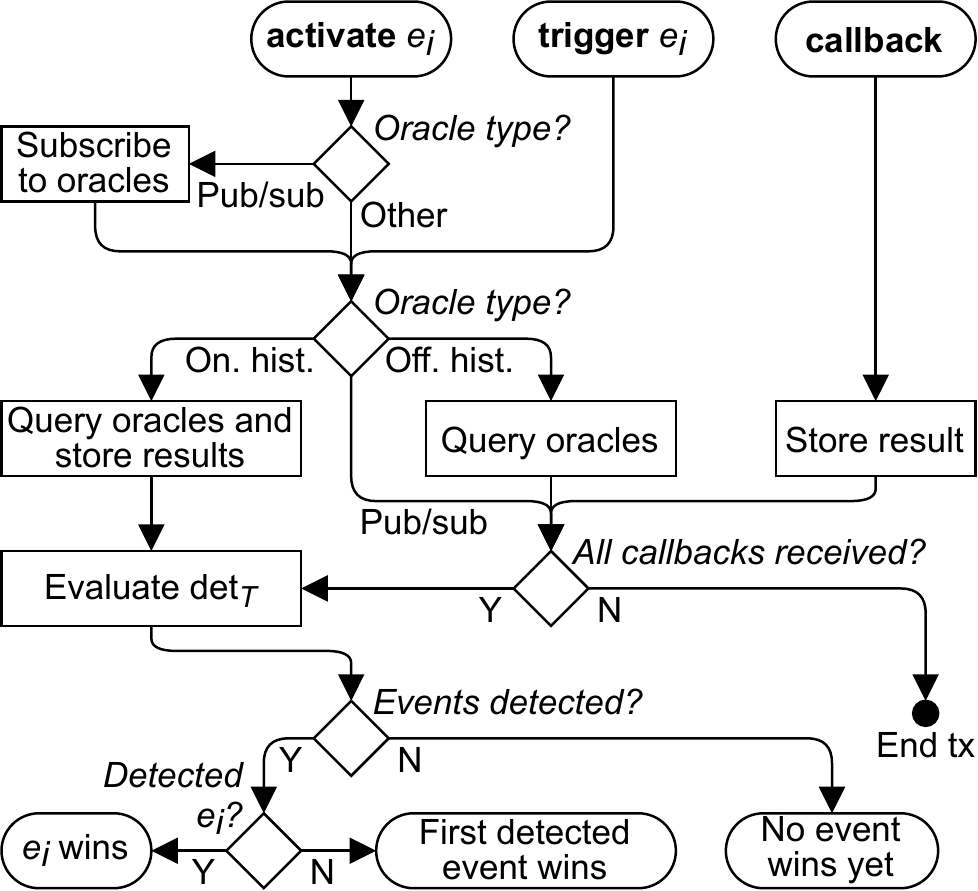}
\caption{Flowchart of the transaction-driven semantics implementation}
\label{fig:contract-semantics}
\end{figure}

Figure~\ref{fig:contract-semantics} shows a schematic overview of the actions performed within all transactions depending on the oracle type.
The complexity of the \async oracles becomes especially apparent since the smart contract needs to keep track of all oracle callbacks, and asynchronously continue the event detection across transactions once the required external variable values are all gathered.

\subsection{Off-Chain Components}
The off-chain components are implemented in Javascript using Node.js and the Ethereum connector library web3.js.

\mysubsubsection{Oracle Providers}
The oracle providers are responsible for bridging the gap between the oracle smart contract and the external variable they observe.
They manage communication responsibilities, mainly updating the oracle smart contract and sending responses to consumer requests depending on the oracle type.
In the scope of this paper, we do not further investigate the connection to the original source of the variable.
Existing approaches using RESTful APIs or similar can be used in practice.

\mysubsubsection{Simulation Framework}
To evaluate our proposals, we implemented a framework to simulate oracles and deferred choices in a reproducible way.
For example, one such simulation may re-enact the scenario used in our running example in Tab.~\ref{tab:example-timelines}.
To this end, a set of \uml{Simulator} instances replay pre-defined lists of actions on their targets, e.g., sending data updates to an oracle provider or calling activation and trigger logic on a deferred choice contract.

\section{Evaluation}
\label{sec:evaluation}

We evaluated our approach using the prototype from two perspectives---the correctness as well as the overall cost.
All simulations were performed on a private Ethereum network with a single node running an official implementation of the Ethereum protocol (Go Ethereum, v1.9.21), using a virtual machine with 12 GB of RAM and 4 CPUs.
The detailed specification and raw results of all simulations are available online alongside the prototype.

\subsection{Correctness}
To verify the correctness of the transaction-driven semantics implementation, we generated $n$ random deferred choices, each with $k$ events $e_0, ..., e_{k-1}$.
An associated simulation timeline was chosen specifically such that the events occur sequentially in order; that is, a transaction is sent for explicit events, a deadline is passed for timer events, or a condition on an oracle becomes true for conditional events.
As a result, event $e_0$ is always the unambiguous winner and must be chosen by a correct approach.

To serve as a baseline, we implemented the regular \storage and \request oracles and the continual semantics as well.
Each of the 10 oracle variants was then used to simulate $n=60$ scenarios, half of which with $k=5$ events and the other half with $k=10$ events.
With a delay of 60~s between subsequent groups of transactions, the experiment took around four days (99:22:10) in total.

\begin{table}
\centering
\small
\caption{Correctness of approaches in $n=60$ random scenarios}
\label{tab:correctness}
\begin{tabular}{l|l||r|r}\hline

\emph{\bfseries Semantics} &
\emph{\bfseries Oracle} &
\multicolumn{2}{l}{\emph{\bfseries Correctness}}
\\\cline{3-4}

&
&
\multicolumn{1}{l|}{\emph{Reg.}} &
\multicolumn{1}{l}{\emph{Cond.}} \\\hline\hline

Continual &
\Storage &
35\% &
35\% \\\cline{2-4}

&
\Req &
35\% &
35\% \\\hline

Transaction- &
On-c. \history &
100\% &
100\% \\\cline{2-4}

driven &
Off-c. \history &
100\% &
100\% \\\cline{2-4}

&
\Pub &
100\% &
100\% \\\hline

\end{tabular}
\end{table}

The share of simulations which yielded the correct winner $e_0$ is shown in Tab.~\ref{tab:correctness} for each regular and conditional oracle variant.
The transaction-driven semantics approaches perform without fail, giving evidence that they indeed describe the intended behavior of deferred choice and that the implementation is accurate.
As expected, the continual semantics using the traditional oracles encounter issues when certain event configurations resembling the problematic example in Sect.~\ref{subsec:problem-statement} are generated---they only pick the correct winner in around 35\% of cases, i.e., those in which the first event randomly turns out be explicit.

\subsection{Cost}
Cost is a major factor that influences the adoption of any approach in practice.
On Ethereum, cost is expressed using \emph{gas}, a stable measure that quantifies the computational complexity and storage requirements of a transaction, and directly translates to the cost in cryptocurrency.
We compare the gas cost of all approaches using a series of simulations.
Again, the traditional oracle patterns \storage and \request are included for comparison.

\mysubsubsection{Simulation Design}
All simulation scenarios follow the same pattern, in which $c$ deferred choices consisting of exactly one conditional event each access a single shared oracle.
The oracle receives $u$ data updates.
A \uml{trigger} transaction is sent to each deferred choice at each fifth update, of which only the last will lead to the conditional event's detection by design.
This recreates a realistic timeline of events for an oracle with multiple consumer contracts.

All simulations were executed sequentially for each oracle variant and for all combinations of $c\in\{ 5,10,20 \}$ and $u\in\{ 1,10,20,30 \}$.
Independent sets of transactions were spaced 40~s apart.
The experiment took a total time of 25:22:33 to finish.

\mysubsubsection{Deployment Cost}
Initially, the smart contracts need to be created on the blockchain, incurring a one-time deployment cost contingent on the code size.
Table~\ref{tab:deployment} shows the average deployment costs we have observed.
For oracles, there are three significant outliers owing to their more complex code: the regular on-chain \history oracle contains code to return the correct slice of historical data, and the two \sync conditional oracles contain code to evaluate conditions.
As expected, they are thus more expensive to deploy.

\begin{table}
\centering
\small
\caption{Average smart contract deployment cost}
\label{tab:deployment}
\newcommand{\perc}[1]{\scriptsize +#1\%}
\begin{tabular}{l||r@{~{\scriptsize/}}r|r@{~{\scriptsize/}}r||r@{~{\scriptsize/}}r|r@{~{\scriptsize/}}r}\hline

\emph{\bfseries Oracle} &
\multicolumn{8}{l}{\emph{\bfseries Avg. deployment cost} ($10^3$ gas)}
\\\cline{2-9}

&
\multicolumn{4}{l||}{\emph{Oracles}} &
\multicolumn{4}{l}{\emph{Deferred Choices}}
\\\cline{2-9}

&
\multicolumn{2}{l|}{\emph{Reg.}} &
\multicolumn{2}{l||}{\emph{Cond.}} &
\multicolumn{2}{l|}{\emph{Reg.}} &
\multicolumn{2}{l}{\emph{Cond.}} \\\hline\hline

\Storage &
276 & & 
408 & \perc{48} & 
1431 & \perc{3} & 
1406 & \perc{1} 
\\\hline

\Req &
281 & \perc{2} & 
281 & \perc{2} & 
1502 & \perc{8} & 
1477 & \perc{7} 
\\\hline

On-c. hist. &
467 & \perc{69} & 
552 & \perc{100} & 
1520 & \perc{10} & 
1386 & 
\\\hline

Off-c. hist. &
281 & \perc{2} & 
281 & \perc{2} & 
1592 & \perc{15} & 
1448 & \perc{4} 
\\\hline

\Pub &
281 & \perc{2} & 
281 & \perc{2} & 
1577 & \perc{14} & 
1490 & \perc{7} 
\\\hline 

\end{tabular}
\end{table}

For deferred choice, the differences are less pronounced.
There is a clear indication, though, that the externalization of evaluation logic to the oracle for the conditional variants reduces the code complexity of the choice itself, and that the more powerful approaches like \publish and \history oracles require larger smart contracts.

\begin{figure}
\centering
\input{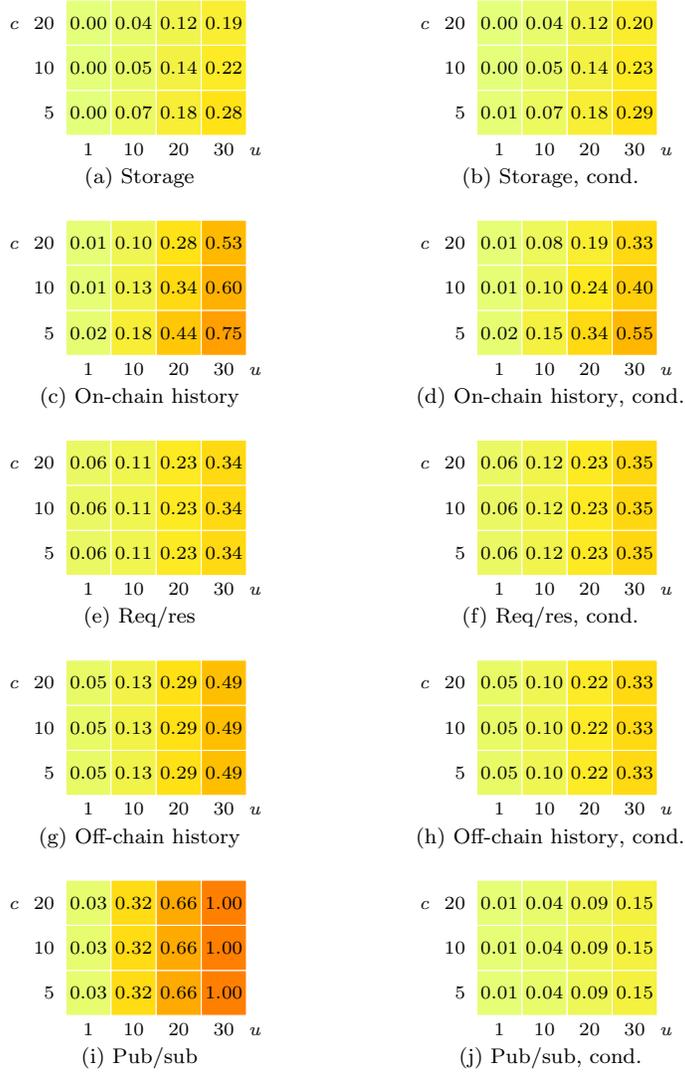}
\caption{Normalized, relative operating cost of an oracle per consumer with the given number of data updates \emph{u} and consumers \emph{c}}
\label{fig:cost}
\end{figure}

\mysubsubsection{Operating Cost}
The operating cost was derived by dividing the total cost minus deployment costs by the number of consumers $c$, arriving at an average cost per consumer.
The results were normalized globally from the minimum (165,407 gas for the \storage oracle with $c=20$, $u=1$) to the maximum (1,656,007 gas for the \publish oracle with $u=30$), producing the overview shown in Fig.~\ref{fig:cost}.

Several observations are immediately apparent:
All \sync oracles (Fig.~\ref{fig:cost}a--d) become relatively less expensive the more consumers share the cost, as visible by the decline along the y-axis.
This is not the case for \async oracles (Fig.~\ref{fig:cost}e--j), as their cost linearly scales with the number of consumers.

Naturally, the more updates there are, the more expensive the approaches tend to get, albeit on different scales.
This is especially evident for the on-chain \history (Fig.~\ref{fig:cost}c--d) and the regular off-chain \history (Fig.~\ref{fig:cost}g) oracles, which show a clear superlinear trajectory on the x-axis because of storage and payload cost increases.

This is not the case for the other oracles, which experience at most a linear growth alongside the number of updates.
Notably, while the regular \publish oracle is the most expensive in our tests, it exhibits a very predictable and linear growth of cost per update, which is not tied to storage or payload requirements.

Interestingly, the conditional variants of the \storage and \request oracles are almost exactly as expensive as their regular counterparts.
This is mainly due to Ethereum's padding of transaction parameters to words (256 bit), making a Boolean value take up just as much space as an integer.
However, for \history oracles the effect is very apparent, and for \publish considerable: the conditional variants outperform the regular variants by a steady margin the larger the payloads get and the more transactions are to be sent.

\section{Related Work}
\label{sec:related}

In this paper, we have considered the semantics of deferred choice in formal terms, proposed concrete oracle architectures, and provided a blockchain-based implementation.
Accordingly, we refer to related work from these three areas.

\mysubsubsection{Deferred Choice Semantics}
In their seminal research on workflow patterns, Russell et al. describe the semantics of deferred choice using colored Petri nets~\cite{russell2006control}.
The ``moment of choice'', that is waiting for an event to occur, is modeled as a place, and each alternative path as an outgoing transition.
Thus, only one branch may be taken per token.
Most subsequent literature to this day adopts these semantics:

Dijkman et al.~\cite{dijkman2008semantics}, for instance, use the Petri net abstraction for analysis and verification purposes.
The \gls{bpmn} standard itself describes a similar token system, but each outgoing branch initially receives a token and all but one are withdrawn after the choice is made~\cite{omg2013bpmn}.
However, none of the works mentioned considers gaps in perception, or specifically discusses the detection of external events.

Some works by Corradini et al.~\cite{corradini2018formal}, Houhou et al.~\cite{houhou2019first}, or Kheldoun et al.~\cite{kheldoun2017formal}, are more focussed on collaborations, and thus spend more time on deferred choice.
However, they only consider explicit message events and their detection.
There is no notion of transactions or external data sources, and no discussion of the interplay with implicit---timer, conditional, or otherwise---events.


Kossak et al. provide the most detailed discussion of the semantics of deferred choice as modeled using event-based gateways in \gls{bpmn}~\cite{kossak2012event}.
Their main focus, however, is on instantiating event-based gateways and their inconsistencies, which are a special flavor of deferred choice that is not immediately applicable to this paper.

Overall, we find that existing literature on the semantics of deferred choice is mostly focused on verification, and if external events are considered at all this is mostly limited to messages.
We thus provide a novel perspective that is especially useful for transaction-driven environments.

\mysubsubsection{Oracle Architectures}
An important aspect of this paper is the proposal of oracle architectures as extended and refined from our own previous work~\cite{ladleif2020external}.
While these are directly based on existing oracle patterns (see Sect.~\ref{subsec:blockchain})~\cite{xu2018pattern}, we extend their capabilities specifically---but not limited---to support of deferred choice.
This focus is, to the best of our knowledge, yet unexplored:

In an exhaustive literature survey focused on trust considerations, Al Breiki et al. identify only three fundamental oracle design patterns in literature and practice~\cite{albreiki2020trustworthy}.
Two of them correspond to the basic \storage and \request patterns.
A third is called \publish pattern by the authors, but is different to the pattern of the same name we propose:
While the goal is to access data ``that is expected to change''~\cite{albreiki2020trustworthy}, this is achieved through on-chain or off-chain flags which are manually polled.
This renders them insufficient for implementing transaction-driven semantics using the \publish strategy (see Sect.~\ref{def:timed-event-detection}), since the active notification of changes is a prerequisite.

Other existing work in the oracle area is chiefly concerned with the core functionality of oracles, that is, providing external data to smart contracts on the blockchain.
As such, trust~\cite{heiss2019oracles}, reliability~\cite{lo2020reliability}, or integration aspects~\cite{mammadzada2020blockchain} are paramount, more so than deviating from standard architectures to support specific workflow patterns.

\mysubsubsection{Deferred Choice on Blockchain}
Numerous approaches at blockchain-based process execution have emerged, ranging from limited prototypes to powerful process engines with support for a wide range of business process constructs (see Sect.~\ref{subsec:blockchain})~\cite{garcia2020blockchain}.
However, deferred choice in its entirety is rarely among those supported constructs.

L{\'o}pez-Pintado et al. implement deferred choice only for internal as well as message events in their Caterpillar engine, which circumvents the problem of gaps in perception since no implicit external events occur~\cite{lopez2019caterpillar}.
Approaches like Lorikeet~\cite{lu2020integrated} and others~\cite{corradini2020engineering,klinger2020cross,lopez2019interpreted,garcia2017optimized,weber2016untrusted} likewise suggest some support for deferred choice for messages and internal events, but do not discuss related issues in detail.
This uncertainty is exacerbated by the fact that prototypical implementations are rarely publicly available.

Still, even though deferred choice does not seem to have been a focus in any existing work, some do consider more types of events.
In our own work, we implement conditional events to monitor local process data~\cite{ladleif2019modeling}.
Some approaches support constructs which can be used to emulate the behavior of external events, like service tasks in Caterpillar~\cite{lopez2019caterpillar} or on-chain asset registries in Lorikeet~\cite{lu2020integrated}.
Weber et al. discuss connecting to external services using a dedicated trigger component, which could also assume the role of a \request oracle~\cite{weber2016untrusted}.
Similarly, many approaches including all of the above allow the inclusion of custom script annotations within their source models, which could potentially be used to access oracles non-natively and emulate resolution of deferred choice (see Fig.~\ref{fig:contract-semantics}) on a process level.

Lastly, temporal constraints and timer events are largely absent in blockchain-based process engines due to the platform's inherent difficulties in these regards~\cite{ladleif2020time}.
They are sometimes mentioned~\cite{ladleif2019modeling,weber2016untrusted,klinger2020cross}, but never discussed in detail.
Abid et al. provide a notable exception and extend Caterpillar with several notions of timer events, albeit not in a deferred choice setting~\cite{abid2020modelling}.

In summary, deferred choice has not been the focus of any approach at blockchain-based process execution yet.
Thus, we find that our work significantly extends on existing research in numerous fields, and is the first to provide an end-to-end vision of the deferred choice pattern on blockchain.

\section{Discussion and Conclusion}
\label{sec:discussion}

Our research has been guided by the research questions postulated in Sect.~\ref{subsec:research-questions}.
On a fundamental level, we are confident that we were able to answer both of them:
We have shown that the semantics of deferred choice can successfully be transferred to transaction-driven environments like the blockchain, and that we can provide feasible implementations in practice.
Some reservations remain, of course:

Perhaps most importantly, we did not take into account some concrete network and protocol delays that are present in blockchains and Ethereum in particular:
There is no notion of forks, side chains, or confirmation time.
As such, our paper presents results obtained in an ideal environment.
At the same time, this largely blockchain-agnostic point of view helps keeping our results generalizable in turn.

Further, the prototype is surely not production-ready.
Authentication and security features were omitted completely, and several optimizations were left out to not unfairly influence the direct comparison of the oracles.
We also do not support some deferred choice and oracle configurations, e.g., multiple types of oracles being used in the same deferred choice instance.
These design decisions were necessary to keep the development scope manageable.

Still, the prototype and results of our evaluation provide important insights for future blockchain-based process engines and associated oracle development.
In particular, it is apparent that supporting deferred choice comes at a price, and the required novel oracle architectures tend to be more expensive than existing approaches with less functionality.
More gravely, due to the platform's storage limitations, some approaches might not be feasible for all data sources and scenarios:

We have shown that on-chain \history oracles might quickly become infeasible when the exponential pricing of storage in Ethereum fully comes into to play with growing storage demands.
There are, of course, conceivable solutions to this, e.g., only storing data for a certain amount of time to keep demands on a constant level.
On the other hand, we have shown that conditional \publish oracles can even outperform existing oracle architectures without any optimization.

To summarize, in this paper we have considered the issues regarding the conceptual as well as practical implementation of the deferred choice workflow pattern in transaction-driven execution environments like the blockchain.
To this end, we formally characterized the source and cause of these issues, and provided several strategies to resolve them.
We adapted blockchain-agnostic oracle architectures to support these strategies in practice, and showed the feasibility of the both the formal and the architectural results by a proof-of-concept implementation.
The fundamental outcomes of this paper may serve as a blueprint for future blockchain-based process engines.

\bibliography{bibliography}

\end{document}